\def\Title{QUANTUM MEASUREMENT, INFORMATION, AND
COMPLETELY POSITIVE MAPS}
\def\Author{Masanao Ozawa}
\documentstyle[12pt]{article}
  \newcommand{\beq}{\begin{equation}}
  \newcommand{\eeq}{\end{equation}}
  \newcommand{\beql}[1]{\begin{equation}\label{eq:#1}}
  \newcommand{\beqa}{\begin{eqnarray}}
  \newcommand{\eeqa}{\end{eqnarray}}
  \newcommand{\beqas}{\begin{eqnarray*}}
  \newcommand{\eeqas}{\end{eqnarray*}}
  \newtheorem{Theorem}{Theorem}
  \newcommand{\R}{{\bf R}}
  
   \newcommand{\bA}{{\bf A}}
   \newcommand{\bL}{{\bf L}}
   \newcommand{\bM}{{\bf M}}
   
   \newcommand{\bS}{{\bf S}}
   
   \newcommand{\bX}{{\bf X}}
  \newcommand{\cA}{{\cal A}}
  \newcommand{\cB}{{\cal B}}
  \newcommand{\cD}{{\cal D}}
  \newcommand{\cE}{{\cal E}}
  \newcommand{\cH}{{\cal H}}
  \newcommand{\cI}{{\cal I}}
  \newcommand{\cK}{{\cal K}}
  \newcommand{\cL}{{\cal L}}
  \newcommand{\cM}{{\cal M}}
  \newcommand{\cS}{{\cal S}}
  \newcommand{\cP}{{\cal P}}
  \newcommand{\cQ}{{\cal Q}}
  \newcommand{\cR}{{\cal R}}
  
  \newcommand{\bx}{{\bf x}}
  \newcommand{\by}{{\bf y}}
  
  \newcommand{\al}{\alpha}

  \newcommand{\da}{\dagger}

  \newcommand{\De}{\Delta}

  \newcommand{\mb}{\mbox}
  \newcommand{\nn}{\nonumber}

  \newcommand{\rh}{\rho}
  \newcommand{\si}{\sigma}

  \newcommand{\vr}{\varrho}
                
  \newcommand{\La}{\Lambda}

  \newcommand{\Ph}{\Phi}

  \newcommand{\Tr}{\mbox{\rm Tr}}

  \newcommand{\tc}{\tau c}
  \newcommand{\eq}[1]{(\ref{eq:#1})}
  \newcommand{\Eq}[1]{(\ref{eq:#1})}
\newcommand{\bra}[1]{\langle#1|}
\newcommand{\ket}[1]{|#1\rangle}
   
\newcommand{\bracket}[1]{\langle#1\rangle}   

\renewcommand{\bX}{\cI}
\renewcommand{\cE}{\cI}
\renewcommand{\bL}{L}

\renewcommand{\Eq}{\eq}
\newcommand{\bvr}{\mbox{\boldmath{$\vr$}}}

  \title{\bf \Title}
  \author{\sc \Author\\
  \small\em  Graduate School of Human Informatics\\ 
  \small\em  Nagoya University, Chikusa-ku, Nagoya 464-8601, Japan}
  \date{}
\begin{document}\maketitle
\abstract{Axiomatic approach to measurement theory is developed.
All the possible statistical properties of apparatuses
measuring an observable with nondegenerate spectrum allowed in 
standard quantum mechanics are characterized.}

\section*{1. INTRODUCTION}

Every measuring apparatus inputs the state $\rh$ of the measured system
and outputs the classical output $\bx$ and the state $\rh_{\{\bx=x\}}$
of the measured system conditional upon the outcome $\bx=x$.
In the conventional approach, the probability distribution of the 
classical output $\bx$ and the output state $\rh_{\{\bx=x\}}$
are determined by the spectral projections of the measured observable 
by the Born statistical formula and the projection postulate, respectively.
This description has been a very familiar principle in quantum mechanics,
but is much more restrictive than what quantum mechanics allows. 
In the modern measurement theory, the problem has been investigated
as to what is the most general description of measurement allowed by 
quantum mechanics.  This paper investigates the problem of the
determination of all the possible measurements of observables with
nondegenerate spectrum and shows that the following conditions 
are equivalent for measurements of nondegenerate 
observables: (i) The joint 
probability distribution of the outcomes of successive measurements depends 
affinely on the initial state. (ii) The apparatus has an indirect 
measurement model. (iii) The state change is described by a positive 
superoperator valued measure. (iv) The state change is described by 
a completely positive superoperator valued measure.  
(v) The family of output states is a Borel family of density operators
independent of the input state and can be arbitrarily chosen 
by the choice of the apparatus. 

\section*{2. MEASUREMENT SCHEMES}

Let $\cH$ be a separable Hilbert space.
%A {\em density operator} on $\cH$ is a positive operator on $\cH$ with 
%unit trace.
The {\em state space} of $\cH$ is the set $\cS(\cH)$ of density operators
on $\cH$.
In what follows, we shall give a general mathematical formulation for
the statistical properties of measuring apparatuses.
For heuristics, we shall consider a measuring apparatus which measures 
the quantum system $\bS$ described by the Hilbert space $\cH$.
Every measuring apparatus has the output variable that gives the outcome 
on each measurement.  We assume that the output
variable takes values in a standard Borel space which is specified by
each measuring apparatus.  We shall denote by $\bA(\bx)$ the  measuring
apparatus with the output variable $\bx$ taking values in a standard
Borel space $\La$ with the Borel $\si$-field $\cB(\La)$.
The statistical property of the apparatus $\bA(\bx)$ consists of
the output distribution $\Pr\{\bx\in\De\|\rh\}$ and
the state reduction $\rh\mapsto\rh_{\{\bx=x\}}$.
The output distribution $\Pr\{\bx\in\De\|\rh\}$ 
describes the probability distribution of the output variable $\bx$
when the input state is $\rh\in\cS(\cH)$, where $\De\in\cB(\La)$.
The state reduction $\rh\mapsto\rh_{\{bx=x\}}$ describes the state
change from the input state $\rh$ to the output state $\rh_{\{\bx=x\}}$,
when the measurement leads to the output $\bx=x$.  
%Since the output
%variable can have a continuous probability distribution, 
The output
state $\rh_{\{\bx=x\}}$ is determined up to probability one with
respect to the output distribution.  The state reduction determines
the collective state reduction $\rh\mapsto\rh_{\{\bx\in\De\}}$ that
describes the output state $\rh_{\{\bx\in\De\}}$ given that  
the output of the measurement is in a Borel set $\De$.  The collective
state reduction is naturally related to the state reduction by the
integral formula
$$
\rh_{\{\bx\in\De\}}
=\frac{1}{\Pr\{\bx\in\De\|\rh\}}
\int_{\De}\rh_{\{\bx=x\}}\Pr\{\bx\in dx\|\rh\}.
$$ 
Formal description of the statistical properties of measuring
apparatuses will be given as follows.

Let $\La$ be a standard Borel space
with $\si$-field $\cB(\La)$.
Denote by $B(\La,\cS(\cH))$ the space of Borel families of
states for $(\cH,\La)$.
The {\em state space} of $\La$ is the set $\cS(\La)$ of probability
measures on $\cB(\La)$.
A {\em measurement scheme} for $(\cH,\La)$ is the pair $(\cP,\cQ)$ of
a function $\cP$ from $\cS(\cH)$ to $\cS(\La)$ and a function $\cQ$ 
from $\cS(\cH)$ to $B(\La,\cS(\cH))$.
The function $\cP$ is called the {\em output probability scheme}, 
and $\cQ$ is called the {\em state reduction scheme}.
Two measurement schemes $(\cP,\cQ)$ and $(\cP',\cQ')$ are said to be
{\em equivalent}, in symbols $(\cP,\cQ)\cong(\cP',\cQ')$, if $\cP=\cP'$ and
$\cQ\rh$ and $\cQ'\rh$
differ only on a null set of the probability measure $\cP\rh$, i.e.,
$$
(\cP\rh)\{x\in\La|\ (\cQ\rh)(x)\not=(\cQ'\rh)(x)\}=0,
$$
for all $\rh\in\cS(\cH)$.

The set of equivalence classes of measurement schemes for $(\cH,\La)$ 
is denoted by $\cM(\cH,\La)$ and 
we define $\cM(\cH)=\bigcup_{\La}\cM(\cH,\La)$.
A {\em measurement theory} for $\cH$ is a pair $(\cA,\bM)$ consisting 
of a nonempty set $\cA$ and a function $\bM$ from $\cA$ to $\cM(\cH)$.
An element of $\cA$ is called an {\em apparatus}.  Every apparatus
has its distinctive {\em output variable}.
We denote the apparatuses with output variables $\bx,\by,\ldots$ by
$\bA(\bx),\bA(\by),\ldots$, respectively. 
We assume that $\bx=\by$ if and only if $\bA(\bx)=\bA(\by)$.
The image of $\bA(\bx)$ by $\bM$ is denoted by $\bM(\bx)$ instead
of $\bM(\bA(\bx))$ for simplicity,
so that $\bM(\bx)$ denotes the equivalence class of the measurement 
scheme corresponding to the apparatus $\bA(\bx)$.
The apparatus $\bA(\bx)$ is called
{\em $\La$-valued} if $\bM(\bx)\in\cM(\cH,\La)$.
For any $\bA(\bx)\in\cA$, the equivalence class $\bM(\bx)$ of 
the measurement scheme is called the {\em statistical property 
of $\bA(\bx)$}.
We shall denote by $(\cP_{\bx},\cQ_{\bx})$ a representative of 
$\bM(\bx)$; in this case, we shall also write 
$\bM(\bx)=[\cP_{\bx},\cQ_{\bx}]$.
The function $\cP_{\bx}$ is called the {\em output probability} 
of $\bA(\bx)$ and $\cQ_{\bx}$ the {\em state reduction} of $\bA(\bx)$
which is determined uniquely up to the output probability one. 
Two apparatuses $\bA(\bx)$ and $\bA(\by)$ 
are said to be {\em statistically equivalent}, 
in symbols $\bA(\bx)\cong\bA(\by)$,
if they have the same statistical property, i.e., $\bM(\bx)=\bM(\by)$.

Suppose that a measurement theory $(\cA,\bM)$ is given.
Let $\bA(\bx)\in\cA$ and $\bM(\bx)=[\cP_{\bx},\cQ_{\bx}]$.
The {\em probability distribution} of the output variable $\bx$
in the state $\rh$ is defined by
$$
\Pr\{\bx\in\De\|\rh\}=(\cP_{\bx}\rh)(\De)
$$
for all $\rh\in\cS(\cH)$ and 
$\De\in\cB(\La)$.  This probability distribution
is called the {\em output distribution} of $\bA(\bx)$ in $\rh$.
The {\em output states} $\{\rh_{\{\bx=x\}}\}_{x\in\La}$
of $\bA(\bx)$ in $\rh$ is defined by
$$
\rh_{\{\bx=x\}}=(\cQ_{\bx}\rh)(x)
$$
for all $\rh\in\cS(\cH)$ and $x\in\La$.

Let $\La_{1},\ldots,\La_{n}$ be standard Borel spaces.
For $j=1,\ldots,n$, let $\bA(\bx_{j})$ be a 
$\La_{j}$-valued apparatus.
A {\em successive measurement in the input state $\rh$} is a sequence of 
measurements using $\bA(\bx_{1}),\ldots,\bA(\bx_{n})$ such that
the input state of the apparatus $\bA(\bx_{1})$ is $\rh$ and
the input state of the apparatus $\bA(\bx_{j+1})$ is the output
state of the apparatus $\bA(\bx_{j})$ for $j=1,\ldots,n-1$.
The joint probability distribution of the outcomes of the 
successive measurements using $\bA(\bx_{1}),\ldots,\bA(\bx_{n})$
in the input state $\rh$ is naturally defined recursively by
\beqa\label{0321a}
\lefteqn{
\Pr\{\bx_{1}\in\De_{1},\ldots,\bx_{n}\in\De_{n}\|\rh\}
}\quad\nn\\
&=&
\int_{\De_{1}}
\Pr\{\bx_{2}\in\De_{2},\ldots,\bx_{n}\in\De_{n}\|\rh_{\{\bx_{1}=x_{1}\}}\}
\Pr\{\bx_{1}\in dx_{1}\|\rh\}
\eeqa
for $\De_{1}\in\cB(\La_{1}),\ldots,\De_{n}\in\cB(\La_{n})$.

Now, we consider the following requirement for a measurement theory
$(\cA,\bM)$:
\vskip\topsep

{\bf Mixing law of the $n$-joint probability distributions ($n$MLPD):}
{\em For any sequence $\bA(\bx_{1}),\ldots,\bA(\bx_{n})$ 
of apparatuses with values in $\La_{1},\ldots,\La_{n}$, respectively,
if the input state $\rh$ is the mixture of $\rh_{1}$ and $\rh_{2}$
such that $\rh=\al\rh_{1}+(1-\al)\rh_{2}$ with $0<\al<1$
then we have
\beqa
\lefteqn{
\Pr\{\bx_{1}\in\De_{1},\ldots,\bx_{n}\in\De_{n}\|\rh\}
}\quad\nn\\
&=&
\al\Pr\{\bx_{1}\in\De_{1},\ldots,\bx_{n}\in\De_{n}\|\rh_{1}\}\nn\\
& &\mb{}+(1-\al)\Pr\{\bx_{1}\in\De_{1},\ldots,\bx_{n}\in\De_{n}\|\rh_{2}\}
\label{eq:cc'}
\eeqa
for all $\rh\in\cS(\cH)$ and 
$\De_{1}\in\cB(\La_{1}),\ldots,\De_{n}\in\cB(\La_{n})$.}
\vskip\topsep

An {\em observable} of $\cH$ is a self-adjoint operator (densely
defined) on $\cH$.
We denote by $E^{A}$ the spectral measure of an observable $A$.
According to the Born statistical formula, 
we say that an $\R$-valued apparatus $\bA(\bx)$ {\em measures an observable
$A$} if
\beql{0603a1}
\Pr\{\bx\in\De\|\rh\}=\Tr[E^{A}(\De)\rh]
\eeq
for all $\rh\in\cS(\cH)$ and $\De\in\cB(\R)$, where $\R$ stands for the
real number field.
A measurement theory $(\cA,\bM)$ is called {\em nonsuperselective}
if for any observable $A$ there is at least one apparatus measuring $A$.

A {\em $\La$-valued observable} of $\cH$ is a projection valued 
measure $E$ from $\cB(\La)$ to $\cL(\cH)$ such that $E(\La)=I$.
The Born statistical formula is generalized as follows.
We say that a $\La$-valued apparatus $\bA(\bx)$ {\em measures a 
$\La$-valued observable $E$} if
\beql{0603a2}
\Pr\{\bx\in\De\|\rh\}=\Tr[E(\De)\rh]
\eeq
for all $\rh\in\cS(\cH)$ and $\De\in\cB(\La)$.
A {\em probability operator valued measure (POVM)} for $(\cH,\La)$
is a positive operator valued measure $F$ from $\cB(\La)$ to 
$\cL(\cH)$ such that $F(\La)=I$.
We say that a $\La$-valued apparatus $\bA(\bx)$ {\em measures a POVM 
$F$} for $(\cH,\La)$ if 
\beql{0603a3}
\Pr\{\bx\in\De\|\rh\}=\Tr[F(\De)\rh]
\eeq
for all $\rh\in\cS(\cH)$ and $\De\in\cB(\La)$.
Conventional measurement theory is devoted to measurements of 
observables but modern theory extends the notion of measurements
to measurements of POVMs [1--5].
We shall describe in the following requirement the essential feature 
of the modern approach.
\vskip\topsep

{\bf Existence of probability operator valued measures (EPOVM):}
{\em For any apparatus $\bA(\bx)$, there exists a POVM $F_{\bx}$ 
uniquely such that $\bA(\bx)$ measures $F_{\bx}$.}
\vskip\topsep

The EPOVM is justified by the following theorem proved in [3].

\begin{Theorem}
For any measurement theory $(\cA,\bM)$, the EPOVM is equivalent to
the 1MLPD.
\end{Theorem}

\section*{3. COLLECTIVE MEASUREMENT SCHEMES}

In order to provide an alternative definition of measurement
schemes, we call a pair $(\cP,\cR)$ as a 
{\em collective measurement scheme} if $\cP$ is 
a function from $\cS(\cH)$ to $\cS(\La)$ and $\cR$ is a function
from $\cB(\La)\times\cS(\cH)$ to $\cS(\cH)$ satisfying
\beql{0317d}
\sum_{n}(\cP\rh)(\De_{n})\cR(\De_{n},\rh)=\cR(\La,\rh)
\eeq
for any countable Borel partition 
$\{\De_{1},\De_{2},\ldots\}$ of $\La$ and $\rh\in\cS(\cH)$,
where the sum is convergent in the trace norm.
The function $\cR$ is called the {\em collective reduction scheme}.
Two collective measurement schemes $(\cP,\cR)$ and $(\cP',\cR')$ are said to be
{\em equivalent}, in symbols $(\cP,\cR)\cong(\cP',\cR')$, if $\cP=\cP'$ and
$\cR(\De,\rh)=\cR'(\De,\rh)$ for all $\De\in\cB(\La)$ with $(\cP\rh)(\De)>0$.

\begin{Theorem}\label{th:CMS}
The relation
\beql{0317c}
(\cP\rh)(\De)\cR(\De,\rh)=\int_{\De}(\cQ\rh)(x)\,d(\cP\rh)(x)
\eeq
where $\De\in\cB(\La)$ and $\rh\in\cS(\cH)$,
sets up a one-to-one correspondence
between the equivalence classes of measurement schemes $(\cP,\cQ)$ 
and the equivalence classes of collective measurement schemes $(\cP,\cR)$.
\end{Theorem}

Let $(\cP,\cQ)$ be a measurement scheme for $(\cH,\La)$.
The collective measurement scheme $(\cP,\cR)$ defined by \Eq{0317c}
up to equivalence is called the {\em collective measurement scheme
induced by $(\cP,\cQ)$} and the function $\cR$ is called the {\em
collective reduction scheme induced by $(\cP,\cQ)$}.

Let $(\cA,\bM)$ be a measurement theory satisfying the 1MLPD.
For any $\bA(\bx)\in\cA$, define $\cR_{\bx}$ to be the collective
reduction scheme induced by $(\cP_{\bx},\cQ_{\bx})$.
The functions $\cR_{\bx}$ is called 
the {\em collective reduction} of the apparatus $\bA(\bx)$.
The {\em collective output states}
$\{\rh_{\{\bx\in\De\}}\}_{\De\in\cB(\La)}$ of $\bA(\bx)$ in $\rh$
is defined by
$$
\rh_{\{\bx\in\De\}}=\cR_{\bx}(\De,\rh)
$$
for all $\rh\in\cS(\cH)$ and $\De\in\cB(\La)$.

\section*{4. DAVIES-LEWIS POSTULATE}

In what follows, we shall introduce some mathematical terminology
independent of particular measurement theory.
A {\em superoperator} for $\cH$ is a bounded linear transformation
on the space $\tc(\cH)$ of trace class operators on $\cH$.
The {\em dual} of a superoperator $\bL$ is the dual
superoperator $\bL^{*}$ defined by 
$\bracket{\bL^{*}A,\rh}=\bracket{A,\bL\rh}$ for all $A\in\cL(\cH)$
and $\rh\in\tc(\cH)$, where $\bracket{\cdot,\cdot}$ stands for 
the duality pairing defined by  $\bracket{A,\rh}=\Tr[A\rh]$ 
for all $A\in\cL(\cH)$ and $\rh\in\tc(\cH)$.
A superoperator  is called {\em positive}
if it maps positive operators to positive operators.
We denote by %$\cL(\tc(\cH))$ the space of superoperators and
$\cP(\tc(\cH))$ the space of positive superoperators.
Positive contractive superoperators are called {\em operations} 
[6].

A {\em positive superoperator valued (PSV) measure}
is a mapping $\cE$ from $\cB(\La)$ to $\cP(\tc(\cH))$
such that if $\{\De_{1},\De_{2},\ldots\}$ is a countable Borel
partition of $\La$, then we have
$$
\cE(\La)\rh=\sum_{n}\cE(\De_{n})\rh
$$
for any $\rh\in\tc(\cH)$, where the sum is convergent in the trace norm.
The PSV measure $\cE$ is said to be {\em normalized}
if it satisfies the further condition  
$$
\Tr[\cE(\La)\rh]=\Tr[\rh]
$$ 
for any $\rh\in\tc(\cH)$.
Normalized PSV measures are called {\em instruments} [2,8]
for short.  

Let $\bX$ be an instrument for $(\La,\cH)$.
The relation
\beql{0604f}
X(\De)=\bX(\De)^{*}I
\eeq
for all $\De\in\cB(\La)$, defines a POVM for $(\cH,\La)$, called the
{\em POVM} of $\bX$.
The relation $T=\bX(\La)$ defines a trace preserving operation, called
the {\em total operation} of $\bX$.

A measurement theory $(\cA,\bM)$ is said to satisfy the 
{\em Davies-Lewis postulate} if it satisfies the follows postulate.
\vskip\topsep

{\bf Davies-Lewis postulate:}
{\em For any apparatus $\bA(\bx)$, there is a normalized PSV measure
$\cE_{\bx}$ satisfying the following relations
for any $\rh\in\cS(\cH)$ and Borel set $\De\in\cB(\La)$:}
\vskip\topsep

(DL1) $\Pr\{\bx\in\De\|\rh\}=\Tr[\cE_{\bx}(\De)\rh]$.
\vskip\topsep

(DL2) ${\displaystyle\rh_{\{\bx\in\De\}}=\frac{\cE_{\bx}(\De)\rh}
                                          {\Tr[\cE_{\bx}(\De)\rh]}}$.
\vskip\topsep

From Theorem \ref{th:CMS}, the normalized PSV measure $\cE_{\bx}$
determines the output state $\rh_{\{\bx=x\}}$ uniquely up to equivalence by
\vskip\topsep

(DL3) ${\displaystyle \int_{\De}\rh_{\{\bx=x\}}\,\Tr[d\cE_{\bx}(x)\rh]
=\cE_{\bx}(\De)\rh}.$
\vskip\topsep

From (DL1), the Davies-Lewis postulate implies 1MLPD.
Although the Davies-Lewis description of measurement is quite general, 
it is not clear by itself whether it is general enough to allow 
all the possible measurements.
The following theorem shows indeed it is the case.

\begin{Theorem}\sloppy
For any nonsuperselective measurement theory,
the Davies-Lewis postulate is equivalent to the 2MLPD.
\end{Theorem}

A measurement theory $(\cA,\bM)$ is called a {\em statistical measurement
theory} if it is nonsuperselective and satisfies 2MLPD.

\begin{Theorem}
Every statistical measurement theory satisfies $n$MLPD for all 
positive integer $n$.
\end{Theorem}

\section*{5. MEASUREMENTS OF OBSERVABLES}

An instrument $\bX$ for $(\La,\cH)$ is said to be {\em decomposable} 
if $\bX(\De)^{*}A=X(\De)T^{*}(A)$ for all $\De\in\cB(\La)$ 
and $A\in\cL(\cH)$, where $X$ is the POVM of $\cI$ and $T$ the total
operation.
For a given $\La$-valued observable $E$ for $(\La,\cH)$, 
an instrument $\bX$ is said to be {\em $E$-compatible} if the POVM of
$\bX$ is $E$. i.e., $\bX^{*}(\De)I=E(\De)$ for all $\De\in\cB(\La)$;
such an instrument is also called {\em observable measuring}.
For an observable $A$, an instrument is called 
{\em $A$-compatible} if it is $E^{A}$-compatible.
The following theorem shows in particular that every observable
measuring instrument is decomposable (see [5, Proposition 4.3] 
for the case of completely positive instruments).

\begin{Theorem}\label{th:DECOMP}
Let $E$ be a $\La$-valued observable of $\cH$.
Let $\bX$ be an $E$-compatible instrument
and $T$ its total operation. Then, we have the following
statements.

{\rm (i)} For any $\De\in\cB(\La)$ and $\rh\in\tc(\cH)$, we have
\beq\label{eq:0117c}
\bX(\De)\rh=T[E(\De)\rh]=T[\rh E(\De)]=T[E(\De)\rh E(\De)].
\eeq

{\rm (ii)} For any $\De\in\cB(\La)$ and $B\in\cL(\cH)$, we have
\beq\label{eq:0117d}
\bX(\De)^{*}B=E(\De)T^{*}(B)=T^{*}(B)E(\De)=E(\De)T^{*}(B)E(\De).
\eeq

%{\rm (iii)} For any $\De\in\cB(\La)$ and $B\in\cL(\cH)$, we have
%\beq\label{eq:0117d'}
%[T^{*}(B),E(\De)]=0.
%\eeq
\end{Theorem}

All the $E$-compatible instruments are determined as follows.

\begin{Theorem}\label{th:A-COMP}
Let $E$ be an $\La$-valued observable of $\cH$.
The relation
\beql{compatible}
\bX(\De)\rh=T[E(\De)\rh]
\eeq
for all $\De\in\cB(\La)$ and $\rh\in\tc(\cH)$
sets up a one-to-one correspondence between 
the $E$-compatible instruments $\bX$
and the $E$-compatible trace preserving operations $T$.
\end{Theorem}

From the above theorem, in every statistical measurement theory
we have the following:
{\em For any apparatus $\bA(\bx)$ measuring a $\La$-valued observable $E$, 
there is an $E$-compatible trace preserving operation $T$ 
such that the statistical property of $\bA(\bx)$ is represented as follows.}
\beqa
\mb{output distribution: }& &
\Pr\{\bx\in\De\|\rh\}=\Tr[E(\De)\rh]\label{eq:0117e}\\
\mb{collective output state: }& &\rh_{\{\bx\in\De\}}
=\frac{T[E(\De)\rh]}{\Tr[E(\De)\rh]}\label{eq:0117f}
\eeqa

\section*{6. MEASUREMENTS OF NONDEGENERATE
OBSERVABLES}

Let $E$ be a $\La$-valued observable for $\cH$.  
We say that $E$ is {\em nondegenerate} if the commutant of $E$ is abelian.
Two Borel families $\{\bvr_{x}\}_{x\in\La}$ 
and $\{\bvr'_{x}\}_{x\in\La}$ of density operators are said to be
{\em $E$-equivalent} if they differ only on an $E$-null set, i.e., 
$$
E\{x\in\La|\ \bvr_{x}\not=\bvr'_{x}\}=0.
$$

\begin{Theorem}\label{th:0315a}\label{th:ND}\sloppy
Let $E$ be a nondegenerate $\La$-valued observable of $\cH$.
The Bochner integral formula
\beql{0312a}
T\rh=\int_{\La} \bvr_{x}\,\Tr[\rh\,dE(x)]
\eeq
for all $\rh\in\tc(\cH)$
sets up a one-to-one correspondence between the $E$-compatible
trace preserving operations $T$
and the $E$-equivalence classes
of the Borel families $\{\bvr_{x}\}_{x\in\La}$
of density operators indexed by $\La$. 
\end{Theorem}

From the above theorem, in any statistical measurement theory
we conclude the following:
{\em For any apparatus $\bA(\bx)$ measuring an $\La$-valued 
observable $E$, 
there is a Borel family $\{\bvr_{x}\}_{x\in\La}$ of density 
operators uniquely up to $E$-equivalence such that the statistical property 
of $\bA(\bx)$ is represented as follows.}
\beqa
\mb{output distribution: }& &\Pr\{\bx\in\De\|\rh\}=\Tr[E(\De)\rh]
\label{eq:nondeg1}\\
\mb{output state: }& &\rh_{\{\bx=x\}}=\bvr_{x}
\label{eq:nondeg2}
\eeqa

\section*{7. INDIRECT MEASUREMENT MODELS}

An {\em indirect measurement model} for $(\La,\cH)$ is 
defined to be a 4-tuple
$(\cK,\si,U,E)$ consisting of a separable Hilbert space $\cK$,
a density operator $\si$ on $\cK$, 
a unitary operator $U$ on $\cH\otimes\cK$,
and a $\La$-valued observable $E$ of $\cK$.

If the apparatus $\bA(\bx)$ has
the indirect measurement model $(\cK,\si,U,E)$, then $\bA(\bx)$
has the instrument $\cE_{\bx}$ defined by
\beql{0604c}
\cE_{\bx}(\De)\rh
=\Tr_{\cK}[(I\otimes E(\De))U(\rh\otimes\si)U^{\da}],
\eeq
where $\De\in\cB(\La)$ and $\rh\in\tc(\cH)$ [10], 
so that the statistical
property of $\bA(\bx)$ is described by $\cE_{\bx}$ with relations
(DL1) and (DL2).  The above instrument $\cE_{\bx}$ is called the
{\em instrument of $\bA(\bx)$}.

Now, we consider the following hypothesis.
\vskip\topsep

{\bf Indirect measurability hypothesis: }
{\em For any indirect measurement model $(\cK,\si,U,E)$, there is an 
apparatus $\bA(\bx)$ with the instrument $\cE_{\bx}$ defined by
\eq{0604c}.}
\vskip\topsep

In general, an instrument $\bX_{\bx}$ is said to 
be {\em realized} by an indirect measurement model $(\cK,\si,U,E)$ if 
\eq{0604c} holds for any $\rh\in\tc(\cH)$.
In this case, the instrument $\cE$ is called {\em unitarily realizable}.
Under the indirect measurability hypothesis, every unitarily realizable
instrument represents the statistical property of an apparatus.

In the sequel, a statistical measurement theory $(\cA,\bM)$
is called a {\em standard measurement theory}, if it satisfies the
indirect measurability hypothesis.
It is natural to consider that any standard measurement theory is
consistent with the standard formulation of quantum mechanics.

\section*{8. COMPLETE POSITIVITY}

Let $\cD=\tc(\cH)$ or $\cD=\cL(\cH)$.
A linear transformation $L$ on $\cD$
is called {\em completely positive (CP)} iff for any finite
sequences of operators $A_{1},\ldots,A_{n}\in\cD$
and vectors $\xi_{1},\ldots,\xi_{n}\in\cH$ we have
$$
\sum_{ij}\bracket{\xi_{i}|L(A_{i}^{\da}A_{j})|\xi_{j}}\ge0.
$$
The above condition is equivalent to that $L\otimes I$
maps positive operators in the algebraic tensor product
$\cD\otimes\cL(\cK)$ to positive operators in 
$\cD\otimes\cL(\cK)$ for any Hilbert space $\cK$.
Obviously, every CP superoperators are positive.
A superoperator is CP if and only if its dual superoperator is CP.
An instrument $\bX$ is called
{\em completely positive (CP)} if every operation $\bX(\De)$ is CP.
It can be seen easily from \eq{0604c} that unitarily
realizable instruments are CP.
Conversely, the following theorem, proved in [5,14], 
asserts that every CP instrument is unitarily realizable.

\begin{Theorem}\label{th:rep}
For any CP instrument $\bX$ for $(\La,\cH)$,
there is a separable Hilbert space $\cK$,
a unit vector $\Ph$ in $\cK$, 
a unitary operator $U$ on $\cH\otimes\cK$,
and a $\La$-valued observable $E$ of $\cK$
satisfying the relation
$$
\bX(\De)\rh
=\Tr_{\cK}[(I\otimes E(\De))U(\rh\otimes\ket{\Ph}\bra{\Ph})U^{\da}]
$$
for all $\De\in\cB(\La)$ and $\rh\in\tc(\cH)$.
\end{Theorem}

The following theorem shows that the complete positivity of observable
measuring instruments is determined by their total operations.

\begin{Theorem}\label{th:A-COMP-CP}
Let $E$ be a $\La$-valued observable.
Then, an $E$-compatible instrument is CP 
if and only if its total operation is CP.
\end{Theorem}

From the above theorem, in any standard measurement theory
we conclude the following statement [5]:
{\em The statistical equivalence classes of apparatuses $\bA(\bx)$ 
measuring a $\La$-valued observable $E$ with indirect measurement 
models are in one-to-one correspondence with the $E$-compatible 
trace preserving CP operations,
where the statistical property is represented by \eq{0117e} 
and \eq{0117f}.}

For the case of nondegenerate observables, we have the following
simple characterizations.

\begin{Theorem}
Let $E$ be a nondegenerate $\La$-valued observable.  Then, every
$E$-compatible operation is completely positive.  
Every $E$-compatible instrument is completely positive.
\end{Theorem}

From the above theorem and Theorem \ref{th:rep},
in the statistical measurement theory we conclude:
{\em Every apparatus measuring a nondegenerate ($\La$-valued) 
observable is statistically equivalent to the one having 
an indirect measurement model.}

\sloppy
Every Borel family $\{\bvr_{x}\}$ of density operators indexed by
$\La$ defines an $E$-compatible trace preserving operation
by Theorem \ref{th:ND},
and it is automatically completely positive so that it is
realized by an indirect measurement model.
Thus, we conclude the following:
{\em   The statistical equivalence classes of apparatuses $\bA(\bx)$ 
measuring a nondegenerate $\La$-valued observable $E$ are in one-to-one
correspondence with the Borel family $\{\bvr_{x}\}$ of density operators
indexed by $\La$,
where the statistical property is represented by \eq{nondeg1} and 
\eq{nondeg2}.}

\end{document}